\documentclass[fleqn,usenatbib]{mnras}
\usepackage{newtxtext,newtxmath}
\usepackage[T1]{fontenc}
\usepackage{ae,aecompl}
\usepackage{graphicx}	
\usepackage{amsmath}	
\newcommand{\psra}{PSR~J1546--5925}
\newcommand{\psrb}{PSR~J1146--6610}
\newcommand{\psrc}{PSR~J0921--5202}

\title[Three millisecond pulsars]{Timing observations of three Galactic millisecond pulsars}
\author[D.R. Lorimer et al.]
{D.R.~Lorimer,$^{1,2}$\thanks{E-mail: duncan.lorimer@mail.wvu.edu (DRL)}
A.M.~Kawash,$^{3}$
P.C.C.~Freire,$^{4}$
D.A.~Smith,$^{5,6}$
M.~Kerr,$^{7}$
\newauthor
M.A.~McLaughlin,$^{1,2}$
M.B.~Mickaliger,$^{8}$ 
R.~Spiewak,$^{8}$ 
M.~Bailes,$^{9,10}$
E.~Barr,$^{4}$ 
\newauthor
M.~Burgay,$^{11}$
A.D.~Cameron,$^{9,10}$
F.~Camilo,$^{12}$
S.~Johnston,$^{13}$
F.~Jankowski,$^{8}$ 
\newauthor
E.F.~Keane$^{14}$,
M.~Keith,$^{8}$
M.~Kramer$^{7}$
and 
A.~Possenti$^{9}$
\\
$^{1}$Department of Physics and Astronomy, West Virginia University, Morgantown, WV 26506-6315, USA\\
$^{2}$Center for Gravitational Waves and Cosmology, Chestnut Ridge Building, Morgantown, WV 26505, USA\\
$^{3}$Center for Data Intensive and Time Domain Astronomy, Department of Physics and Astronomy,\\ Michigan State University, East Lansing, MI 48824, USA\\
$^{4}$Max-Planck-Institut f\"ur Radioastronomie (MPIfR), Auf dem H\"ugel 69, D-53121 Bonn, Germany\\
$^{5}$Centre d'\'Etudes Nucl\'eaires de Bordeaux Gradignan, IN2P3/CNRS, Universit\'e Bordeaux, 33175 Gradignan, France\\
$^{6}$Laboratoire d'Astrophysique de Bordeaux, Universit\'e Bordeaux, B18N, all\'ee Geoffroy Saint-Hilaire, 33615 Pessac, France\\
$^{7}$Space Science Division, Naval Research Laboratory, Washington, DC 20375, USA\\
$^{8}$Jodrell Bank Centre for Astrophysics, University of Manchester, Oxford Road, Manchester M13 9PL, UK\\
$^{9}$Centre for Astrophysics and Supercomputing, Swinburne University of Technology, PO Box 218, Hawthorn, VIC 3122, Australia\\
$^{10}$ARC Centre of Excellence for Gravitational Wave Discovery (Ozgrav)\\
$^{11}$INAF, Osservatorio Astronomico di Cagliari, Via della Scienza 5, I-09047, Selargius (CA), Italy\\
$^{12}$South African Radio Astronomy Observatory, Cape Town, South Africa\\
$^{13}$CSIRO Astronomy and Space Science, Australia Telescope National Facility, PO~Box~76, Epping NSW~1710, Australia\\
$^{14}$National University of Ireland Galway, University Road, Galway, Ireland
H91 TK33}

\pubyear{2021}
\date{Accepted 2021 August 27. Received 2021 August 25; in original form 2021 August 09}
\hypersetup{draft}
\begin{document}
\label{firstpage}
\pagerange{\pageref{firstpage}--\pageref{lastpage}}
\maketitle

\begin{abstract}
We report observed and derived timing parameters for three millisecond pulsars (MSPs) from observations collected with the Parkes 64-m telescope, Murriyang. The pulsars were found during re-processing of archival survey data by Mickaliger et al. One of the new pulsars (\psra) has a spin period $P=7.8$~ms and is isolated. The other two (\psrc\ with $P=9.7$~ms and \psrb\ with $P=3.7$~ms) are in binary systems around low-mass ($>0.2 M_{\odot}$) companions. Their respective orbital periods are $38$.2~d and $62.8$~d. While \psrc\ has a low orbital eccentricity $e=1.3 \times 10^{-5}$, in keeping with many other Galactic MSPs, \psrb\ has a significantly larger eccentricity, $e = 7.4 \times 10^{-3}$. This makes it a likely member of a group of eccentric MSP--He white dwarf binary systems in the Galactic disk whose formation is poorly understood. Two of the pulsars are co-located with previously unidentified point sources discovered with the {\it Fermi} satellite's Large Area Telescope, but no $\gamma$-ray pulsations have been detected, likely due to their low spin-down powers. We  also show that, particularly in terms of orbital diversity, the current sample of MSPs is far from complete and is subject to a number of selection biases.
\end{abstract}

\section{Introduction}\label{sec:intro}

Understanding the demographics, origin and evolution of pulsars depends strongly on having a well-determined sample based on sensitive surveys. The Parkes radio telescope, also known as Murriyang\footnote{The name Murriyang represents the ``Skyworld'' where a prominent creator spirit of the Wiradjuri Dreaming, Biyaami, lives.} in Wiradjuri, has been at the forefront of much of this effort. Since the initiation of the Parkes Multibeam Pulsar Survey \citep{2001MNRAS.328...17M}, this and follow-up efforts using the original analogue data acquisition system led to the discovery of 944 pulsars along the Galactic plane and at intermediate and high latitudes \citep{2001MNRAS.328...17M,2002MNRAS.335..275M,2003MNRAS.342.1299K,2004MNRAS.352.1439H,2004MNRAS.355..147F,2006MNRAS.372..777L,2015MNRAS.450.2185L,2009MNRAS.395..837K,2010MNRAS.407.2443E,2012ApJ...759..127M,2013MNRAS.431..292E,2013ApJ...774...93K}. Statistical analyses of the sample of pulsars from this and other Parkes Multibeam surveys covering parts of the Southern sky have substantially improved our knowledge of the radio pulsar population \citep[see, e.g,][]{2006ApJ...643..332F,2006MNRAS.372..777L}. 

\begin{table*}
\label{tab:timing}
\centering
\caption{Observed and derived parameters of the three MSPs. Timing parameters are specified in Dynamic Barycentric Time (TDB). The orbital parameters were obtained with the ELL1 model (see text for details). Figures in parentheses give the 1-$\sigma$ uncertainties in the least significant digits. The first distance is estimated from the DM using the \citep{cl02} electron density model, and the second uses the \citep{ymw17} electron model. Pseudoluminosities are given as the flux density times distance squared. The minimum companion mass $M_c$ is deduced from the mass function $f=(M_c \sin i)^3 / (M_p + M_c)^2$ assuming a pulsar mass $M_p = 1.4\,M_{\odot}$ and an orbital inclination angle $i =90$ degrees. Limits on the total proper motion were derived from an independent fit.
}
\begin{tabular}{@{}llll}
\hline
Pulsar & \psrc\ & \psrb\ & \psra \\
\hline
\multicolumn{4}{l}{General parameters}\\
\hline
Reference epoch (MJD) & 57041& 57085 &57071\\
Data span (MJD) & 56767--57314 & 56767--59183 & 56767--57374\\
Fit $\chi^2$ / number of degrees of freedom &  11.98/12 & 155.66/156 & 24.76/25 \\
EFAC & 1.16 & 1.26 & 0.96 \\
Post-fit RMS of residuals ($\mu$s) & 28.8 & 53.4 & 19.8\\
Right ascension, $\alpha$ (epoch J2000; hh:mm:ss.s) & 09:21:59.8785(7) &
11:46:58.3731(7) & 15:46:29.3456(5) \\
Declination, $\delta$ (epoch J2000; dd:mm:ss.s) & $-$52:02:38.47(1) &
$-$66:10:51.358(3) & $-$59:25:44.871(3) \\
Total proper motion, (mas yr$^{-1}$) & $< 216 $ & $< 8$ & $< 32$ \\
Spin period, $P$ (ms) & 9.679813614785(8)  & 3.722312502674(2) & 7.796728856036(2) \\
First derivative of spin period, $\dot{P}$ ($10^{-20}$ s s$^{-1}$) 
& 1.7(1) & 0.793(6)  & 1.61(3)\\
Dispersion measure, DM (cm$^{-3}$ pc) & 122.4(6) & 133.82(1) & 168.5(5)\\
Flux density at 1.4 GHz (mJy) & 0.29(2) & 0.55(4) & 0.67(5)\\
\hline
\multicolumn{4}{l}{Orbital parameters}\\
\hline
Orbital period, $P_b$ (d)    & 38.223675(1) & 62.7711887(6) & - \\
Projected semi-major axis, $x$ (lt sec) & 19.05341(2) &  22.858617(7) & - \\
First Laplace-Lagrange parameter, $\epsilon_1$          & $-$0.0000102(14) & 
0.0071059(7) & -  \\
Second Laplace-Lagrange parameter, $\epsilon_2$          & 0.0000096(17) & $-$0.0020616(6) & - \\
Time of ascending node, $T_{\rm asc}$ (MJD)      & 56769.836729(6) & 56750.930167(8) & - \\
\hline
\multicolumn{4}{l}{Derived parameters}\\
\hline
Galactic longitude, $l$ (degrees) & 273.78
    &296.48   &323.58   \\
Galactic latitude, $b$ (degrees) & --1.45
    &--4.12   &--3.77   \\
Distance (NE2001, YMW17) (kpc) & 1.7, 0.4 & 2.8, 1.8 & 3.4, 3.9\\
Pseudoluminosity at 1.4 GHz (mJy~kpc$^2$) & 0.8, 0.04 & 4.3, 1.8 & 7.7, 10.2\\
Spin-down luminosity ($10^{25}$~W) & 7.3 & 61 & 13 \\
Characteristic age (Gy) & 9.1 & 7.4 &7.7\\
Surface magnetic field strength ($10^4$~T) & 4.1 & 1.7 & 3.6\\
Orbital eccentricity, $e$                        & 0.0000140(16) & 0.0073989(7) & - \\
Longitude of periastron, $\omega$                & 313(6) & 106.179(5) & - \\
Time of passage through periastron, $T_0$        & 56803.1(7) & 56769.4440(9) & - \\
Mass function, $f$ ($M_{\odot}$)  & 0.005083183(14)  &  0.003254707(3)  & -  \\
Minimum companion mass, $M_{\rm c}$ ($M_{\odot}$)  & 0.24   & 0.20 & - \\
\hline
\end{tabular}
\end{table*}

Thanks to the Parkes Multibeam surveys and many others over the past two decades \citep[see, e.g.,][]{2010MNRAS.409..619K,2016ApJ...819...34C}, the pace of millisecond pulsar (MSP) discoveries has greatly increased. The current sample of MSPs in the Galactic field, which we here define somewhat arbitrarily as pulsars having periods $P<30$~ms, now exceeds 400. Of these, 30\% were discovered in deep radio searches \citep{Ray2012psc} of {\it Fermi} Large Area Telescope (LAT) $\gamma$-ray sources without known counterparts \citep[as in e.g.][hereafter 4FGL]{4FGL}. This has led to an era in which some of the new discoveries are known to a relatively small subset of the community prior to publication and their subsequent appearance in the pulsar catalog\footnote{\url{https://www.atnf.csiro.au/research/pulsar/psrcat}} \citep{2005AJ....129.1993M}. Since 2012, to broaden the availability of this information prior to full publication, we have maintained an online\footnote{\url{http://astro.phys.wvu.edu/GalacticMSPs}, maintained by Elizabeth C. Ferrara.} list of MSPs in the Galactic disk. For each MSP, we list its name, Galactic coordinates, pulse period, dispersion measure and information on the survey(s) which detected it, as well as year of  discovery. For binaries, we list preliminary measurements of the orbital period and semi-major axis. In addition, FAST \citep[see e.g.][]{pqm+21} and MeerKAT \citep[see e.g.][]{rgf+21} are currently leading a surge in discoveries of MSPs in globular clusters\footnote{\url{http://www.naic.edu/~pfreire/GCpsr.html}}.

In this paper, we describe timing observations for three MSPs found during the reprocessing of the Parkes Multibeam Pulsar Survey by \cite{2012ApJ...759..127M}, named PSRs J0922$-$52, J1147$-$66, and J1546$-$59 in that work. Timing solutions have already been published for the two other MSPs from that search, PSR~J1227$-$6208 \citep{2015MNRAS.446.4019B} and PSR~J1725$-$3853 \citep{2012ApJ...759..127M}. One further MSP found in this analysis, PSR~J1753--2822 \citep{2013PhDT.......402M}, was subsequently timed and described by \citet{2019MNRAS.487.1025P}. In \S~\ref{sec:observations}, we describe the observations and present the new timing solutions. In \S~\ref{sec:discussion}, we discuss specific aspects of these new pulsars and place them in context with what is known about the MSP sample. We present our conclusions and suggestions for future work in \S~\ref{sec:conc}.

\section{Observations}
\label{sec:observations}

\subsection{Radio timing at Parkes}
\label{sec:timing}

Regular observations of the three MSPs have been carried out with the central beam of the Parkes  Multibeam receiver as part of an organised follow-up campaign that involves participants from a number of different search projects. We used the BPSR back-end \citep{Keith10,2011AIPC.1357..351S} to channelise and record the data as 1024 channels over a 400~MHz band sampled with 8-bit precision and recorded to disk as 2-bit integers every 64~$\mu$s, along with some preliminary observations using the now defunct 1-bit analogue filterbank system, with 512 channels spanning a 256 MHz band \citep{2001MNRAS.328...17M} sampled every 80~$\mu$s.

The latter data were used to compile measurements of barycentric spin periods as a function of time from which we were able to determine preliminary ephemerides \citep[as described, e.g., in][]{2004hpa..book.....L} for all three pulsars, including orbital parameters for the two binary systems, \psrc\ and \psrb. These were then used to fold the BPSR observations to produce {\tt psrfits} formatted pulse profiles \citep{2004PASA...21..302H}.
All these data were collected between April 2014 and March 2017. For \psrb, 
we made 3 additional observations in October and November 2020, using the UWL receiver and the Medusa back-end \citep{Hobbs20}.

From each profile, a pulse time of arrival (TOA) was computed by Fourier domain fitting with a noise-free template profile of high signal-to-noise ratio \citep[for details, see][]{1992RSPTA.341..117T}. Due to the generally sparse coverage of the observations at Parkes, the initial orbital solutions for \psrc\ and \psrb\ found above were insufficient to phase-connect these two pulsars using standard bootstrapping pulsar timing techniques. To address this, we made use of a new technique recently developed and described in detail by \cite{2018MNRAS.476.4794F} in which offsets between groups of TOAs are initially used to refine preliminary ephemerides which are then iteratively examined for pulse numbering ambiguities between epochs to  yield a full phase-coherent timing solution\footnote{\url{https://github.com/pfreire163/Dracula}}. The resulting timing residuals are free from systematic trends. 

Table 1 summarizes the spin, astrometric, and orbital parameters. These result from an analysis of the TOAs made using the TEMPO timing software\footnote{\url{https://sourceforge.net/projects/tempo/}}, where we used the DE440 Solar System ephemeris \citep{2021AJ....161..105P} and the Bureau International des Poids et Mesures BIPM2019 time scale. The units are quoted in Dynamic Barycentric Time (TDB). As usual in pulsar timing, we have increased the quoted TOA uncertainties by a factor that is listed in the table as EFAC, in order to obtain a reduced $\chi^2$ of the TOA residuals of 1.0; this results in more realistic uncertainty estimates. The latter are 1-$\sigma$ equivalent uncertainties, they are indicated in parentheses, and apply to the last digit of their corresponding value. The spin period is quoted for the reference epoch. To minimize correlations between parameters, this epoch is chosen to be near the centre of each data span. The exception is \psrb, where it denotes the centre of our earlier data set, without the late 2020 TOAs. The proper motion has not been detected for any of these pulsars; for this reason we derive only upper limits on the total proper motion, which are also presented in Table 1. Hence no corrections to the spindown rate were applied. For a typical transverse MSP velocity of 100 km s$^{-1}$ and the \citep{cl02} dispersion measure (DM) distances in Table 1, the corrected $\dot P$ values would be between 67\% and 31\% of those observed. The expected contributions from Galactic acceleration for these pulsars are much smaller. Using Eqs.~16 and 17 from \citet{2009MNRAS.400..805L}, we estimate $\dot{P}$ contamination due to this effect to be at the level of (3--9)$\times 10^{-22}$.

The orbital parameters for the two binaries (\psrc\, and \psrb) were derived using the ELL1 model \citep{2001MNRAS.326..274L}.
When the orbital eccentricity, $e$, is small, the longitude of periastron ($\omega$) is not precisely determined, and it becomes very highly correlated with the time of passage through periastron ($T_0$); the same happens for $P_b$ and $\dot{\omega}$. The ELL1 model addresses this by referring instead to the time of passage through ascending node, $T_{\rm asc}$, which can be defined precisely even for circular binaries, and using the two Laplace-Lagrange parameters, $\epsilon_1 = e \sin \omega$ and $\epsilon_2 = e \cos \omega$. The downside is that the equations are not exact, but a sum of terms of order $x e^{n}$, where $x$ is the projected semi-major axis and $n$ is an integer. The TEMPO and TEMPO2 implementations of this model have recently been updated with terms of order $xe^2$ \citep{2019ApJ...881..165Z}; this implies that this model can now describe, with great accuracy, most MSP - WD binaries known. This is certainly the case for \psrc. For \psrb, the magnitude of $e$ implies that the model is borderline applicable: the largest neglected terms, of order $xe^3$, have a magnitude of $9.3 \, \mu$s, which is of the same order of magnitude as the error in $x$, $7.5 \, \mu$s. Future, more precise timing with MeerKAT \citep{MeerKATpsrs,2021MNRAS.504.2094K} will require an exact description, like the DD model \citep{1986AIHS...44..263D}, especially if we aim at measuring the Shapiro delay for this system.

\begin{figure}
\includegraphics[width=0.47\textwidth]{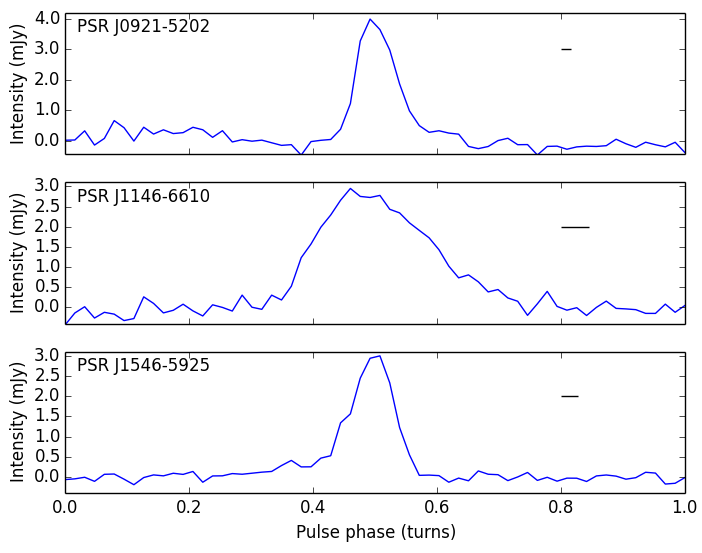}
\caption{Integrated pulse profiles for the three MSPs. The horizontal lines at phase 0.8 show the effective time resolution of each profile computed from a quadrature sum of the sampling time and dispersion measure smearing (i.e., neglecting the effects of multi-path scattering).}
\label{fig:profs}
\end{figure}

In addition to the timing parameters in Table 1, we also provide $1.4$-GHz flux density measurements obtained from an analysis of the pulse profiles from each observation. In the absence of a noise diode calibration scheme, for each folded  pulse profile, we compute the off-pulse root-mean-square value and then scale the profile so that it matches the expected number in janskys computed from radiometer noise considerations \citep[for details, see Chapter 7 of][]{2004hpa..book.....L}. Also taken into account in this calculation is the telescope's offset from the true position of the pulsar found from the timing analysis described above. We assume a Gaussian beam with full-width half maximum of $14'$~\citep{2001MNRAS.328...17M} and scale each profile by the inverse of the degradation factor caused by the offset pointing. As a final step, we integrated the pulse profiles for the three MSPs by summing all the observations, using the timing solutions to achieve phase alignment, as seen in Fig.~\ref{fig:profs}.

\subsection{High-energy follow up with the {\it Fermi} LAT}

Given the precise positions for these three MSPs, we have looked for their counterparts at other wavelengths. The presumed helium white dwarf (He WD) companions of \psrc\, and \psrb\, can in principle be detected in the optical and near IR bands, but no sources at the positions of these pulsars were detected in either the DECam plane survey \citep{2018ApJS..234...39S}, the deepest optical survey covering the search area, or the 2MASS survey \citep{2006AJ....131.1163S}; and none has a position coincident with any source in the 3$^{\rm rd}$ GAIA data release \citep{2021A&A...649A...1G}. This is not surprising given the large distances and low Galactic latitudes, this means that any companions will likely be significantly reddened by intervening dust.

In what follows, we will elaborate on the search for $\gamma$-ray counterparts. Since its launch in 2008, the Large Area Telescope (LAT) on the {\it Fermi} satellite has detected GeV $\gamma$-rays from nearly 300 pulsars, about 250 of which are published at the timing of writing\footnote{\url{https://confluence.slac.stanford.edu/display/GLAMCOG/Public+List+of+LAT-Detected+Gamma-Ray+Pulsars}}. The first 117 were characterised by \citet{2013ApJS..208...17A}, and the rest will be in the third $\gamma$-ray pulsar catalog (in preparation). Half of the pulsars in this sample are MSPs. Steady counterparts for 290 pulsars are in the 4FGL catalog (Data Release 3). Most known $\gamma$-ray MSPs are within 4 kpc of Earth and have spin-down power $10^{26} < \dot E < 10^{28}$~W. The fraction of known MSPs detected in $\gamma$-rays rises from $<40$\% to $>75$\% over this $\dot E$ range \citep[][]{LaffonNewPSRs}. 

While all three MSPs presented here have distances like currently known $\gamma$-ray MSPs, $\dot E$ values for \psrc\ and \psra\ lie at the low end of the typical range. \psrc\ is within the large ($0.5^\circ$) error ellipse of the faint ($5.8 \sigma$) source 4FGL J0924.1$-$5202, while \psra\ has no putative counterpart. \psrb\ has a typical $\gamma$-ray MSP $\dot E$ and distance, and lies at the edge of the error ellipse for 4FGL J1147.7$-$6618 ($6.5 \sigma$ steady point source detection). If the two LAT sources correspond to the two pulsars, the $\gamma$-ray efficiency $\epsilon = L_\gamma/\dot E$ would be near 20\% for both, with $L_\gamma$ the luminosity above 100 MeV for the \citet{ymw17} DM distances. The $\gamma$-ray MSP population covers all latitudes, but these three MSPs have Galactic latitude $-4.1^\circ < b < -1.4^\circ$, where source confusion is high. LAT sensitivity at the pulsar positions, as well as the measured integral energy flux of the two co-located sources, is in the range 3--$9 \times 10^{-15}$~W~m$^{-2}$. 

We first searched for $\gamma$-ray pulsations from the three MSPs following the method of \citet{ThousandFold}. We selected $\gamma$-ray events with energy above 100 MeV within $2^\circ$ of the timing position. For each event, we calculated the rotational phase corresponding to the photon arrival time using the ephemerides presented in this paper and the {\it fermi} plugin to \textsc{Tempo2} \citep{Hobbs2006}. We { weighted} the photons using the simple prescription in  \citet{ThousandFold}, which involves calculating the probability that a $\gamma$-ray of that energy coming from the pulsar could have its angular distance from the pulsar position, given the LAT's energy-dependent angular resolution. For \psrc\ and \psra, the statistical significance of a deviation from a uniform distribution in the resulting phase histograms was $<1\sigma$, that is, no evidence for pulsations, both over the $\sim 1.5$-year ephemeris validity period and over the 12-year LAT data sample. The spectral energy distributions of both co-located sources show significant pulsar-like curvature.

Given its spin-down power and co-location with a 4FGL source, \psrb\, is the best candidate among the three MSPs for $\gamma$-ray counterpart. We searched for $\gamma$-ray pulsations using the more sensitive method of \citet{SearchPulsation}. For this, we descended to $50$~MeV and extended to $5^\circ$. For both weighting methods, the significance of a putative deviation from a uniform phase distribution was $2.1\sigma$ for the full data set, and $2.4\sigma$ during the 6.6~yr of ephemeris validity. \citet{ThousandFold} showed that in such analyses $4\sigma$ significances are required to avoid false positive detections. If the deviation is due to a weak pulsed signal, continued radio timing would provide the phase-connected ephemeris necessary to phase-fold the gamma rays, and a $4\sigma$ pulsed gamma-ray detection could appear around 2025.

\section{Discussion}
\label{sec:discussion}

\subsection{Two binary MSPs}

Assuming pulsar masses $M_p = 1.4 \, M_{\odot}$, the two new binary MSPs presented in this work, \psrc\, and \psrb, have minimum companion masses $M_c > 0.24$ and $>0.20 \, M_{\odot}$. For their orbital periods, \cite{1999A&A...350..928T} predict He WD masses of $\sim$ 0.30 and $\sim\, 0.32 \, M_{\odot}$, thus compatible with the minimum values of both binaries. If true, the latter values imply that both systems are seen at relatively small orbital inclinations: for the assumed pulsar mass these are about $54^\circ$ and $41^\circ$, respectively. The pulsars themselves are fast-spinning and have very small magnetic fields (see Table 1), in agreement with what is commonly observed among MSP--He WD systems with these orbital periods. The orbital eccentricity of the \psrc\, system is also in rough agreement with the predictions of \cite{1992RSPTA.341...39P} for its orbital period. However, this is not the case for the \psrb\, system, where $e$ is about two orders of magnitude larger than that prediction. We discuss this in more detail below.

\begin{table*}
\centering
\caption{Parameters for the eccentric binary MSPs known in the Galactic disk, ordered by increasing orbital period. This selection excludes the currently known double neutron star binary systems. Note how PSR~J1903+0327, which has a confirmed main-sequence star companion, differs from the remaining systems, for which the companions are thought to be He WDs; this has been confirmed for the companion of PSR~J2234+0611 \citep{2016ApJ...830...36A}. Although \psrb\, belongs to the eccentric MSP--He WD systems, it has a few important differences from the others. For PSR~J1903+0327, the uncertainties on the masses are strongly non-Gaussian, they are presented here to 99.7\% confidence limit. For systems where there are no published masses yet, we calculate a minimum companion mass assuming the pulsar mass in square brackets.\label{tab:eMSPs}}
\begin{tabular}{r r l l l l c}
\hline
\multicolumn{1}{c}{PSR} & \multicolumn{1}{c}{$P$} & \multicolumn{1}{c}{$P_b$} & \multicolumn{1}{c}{$e$} & \multicolumn{1}{c}{$M_p$} & \multicolumn{1}{c}{$M_c$} & References \\
       & \multicolumn{1}{c}{(ms)} & \multicolumn{1}{c}{(d)}  &     & \multicolumn{1}{c}{($M_{\odot}$)} & \multicolumn{1}{c}{($M_{\odot}$)} \\
\hline
 J1950+2414   & 4.3048 & 22.1914 & 0.0798  & 1.496(23) & $0.280^{+0.005}_{-0.005}$ & \cite{2015ApJ...806..140K,2019ApJ...881..165Z} \\
 J1618$-$3921 & 11.9873 & 22.7456 & 0.0274 & [1.4] & $> 0.18$ & \cite{2001ApJ...553..801E,2018AA...612A..78O} \\
 J0955$-$6150 & 1.9993 & 24.5784 & 0.1175  & [1.4] & $> 0.22$ & \cite{2015ApJ...810...85C} \\
 J1946+3417   & 3.1701 & 27.0199 & 0.1345  & 1.828(22) & 0.2556(19) & \cite{2013MNRAS.435.2234B,2017MNRAS.465.1711B} \\
 J2234+0611   & 3.5766 & 32.0014 & 0.1293  & $1.353^{+0.014}_{-0.017}$ & $0.298^{+0.015}_{-0.012}$ & \cite{2013ApJ...775...51D,2019ApJ...870...74S} \\
 J1146--6610 & 3.7223 & 62.7712 & 0.0074  & [1.4] & $> 0.20$ & This paper \\
\hline
 J1903+0327   & 2.1499 & 95.1741 & 0.4367 & 1.667(21) & 1.029(8) & \cite{2008Sci...320.1309C,2011MNRAS.412.2763F} \\
\hline
\end{tabular}
\end{table*}
\subsection{Eccentric binary MSPs in the Galactic disk}

One of the key predictions of binary stellar evolution theory is that, after recycling of a neutron star via accretion of matter from a stellar companion, the orbit will be circularised by tidal interactions (e.g., \citealt{1982Natur.300..728A}). The system can become eccentric again if the companion undergoes a supernova (e.g., \citealt{2017ApJ...846..170T}), but if it evolves instead into a WD, then the orbit should retain a very low eccentricity \citep{1992RSPTA.341...39P}. This is indeed what is observed for the vast majority of recycled pulsars outside globular clusters. In globular clusters,  MSPs with high eccentricities arise due to the large stellar densities which can significantly perturb the orbits of pulsar--WD binaries after the recycling (\citealt{1992RSPTA.341...39P} and references therein); they can even cause exchanges of companions (e.g., \citealt{1991ApJ...374L..41P}). 

However, there are exceptions, which are listed in Table~\ref{tab:eMSPs}. In 2008, a highly eccentric  binary MSP, PSR~J1903+0327, was discovered in the Galactic disk \citep{2008Sci...320.1309C}. This did not originate in a globular cluster or the Galactic centre, and has a 1~$M_{\odot}$ main sequence star as a companion \citep{2011MNRAS.412.2763F}. The most likely hypothesis is that this system formed via the chaotic disruption of a hierarchical triple system \citep{2011MNRAS.412.2763F,2011ApJ...734...55P}. 

Following this, an additional five moderately eccentric ($0.027 < e < 0.14$) MSP binaries have been previously characterized, these are the first five systems in Table~\ref{tab:eMSPs}. They have very similar orbital characteristics.  Apart from their eccentricities, their orbital periods are all between 22 and 32 d, occupying a previously identified gap in distribution of orbital periods for binary MSPs.  The measured companion masses (see Table~\ref{tab:eMSPs}) are also compatible with the predictions of \cite{1999A&A...350..928T} for He WDs (for PSR~J2234+0611 the companion was confirmed to be a He WD by optical observations, see \citealt{2016ApJ...830...36A}), although for PSR~J1946+3417, the companion mass is somewhat smaller than that expectation.

For \psrb, $e$ is two orders of magnitude larger than predicted for its orbital period by \cite{1992RSPTA.341...39P}. For this reason we consider it the seventh eccentric MSP system known in the Galactic disk. Its companion mass is consistent with that of a He WD, however, its orbital period is $\gtrsim 2$ times larger and $e$ is one order of magnitude smaller than those of the other eccentric MSP--He WD systems, a combination of characteristics that makes \psrb\, unique. In what follows, we will discuss the formation of these enigmatic systems.

\subsection{Formation of eccentric MSPs}

The orbital similarities of the first five systems in Table~\ref{tab:eMSPs} suggest that their origin is a well-defined stellar evolution mechanism with a predictable outcome, unlike the chaotic disruption of a triple system that is thought to have formed PSR~J1903+0327. Two hypothetical mechanisms \citep{2014MNRAS.438L..86F,2015ApJ...807...41J} suggest that these eccentricities result from the release of gravitational binding energy that happens in a phase transition in the object that later becomes the MSP. This necessarily occurs after the mass transfer phase has ended, otherwise the system would re-circularize. This delayed collapse is therefore not caused by mass accretion, it is caused instead, in both hypotheses, by the spin-down of the progenitor to the MSP, which slowly decreases the centrifugal force, and thus leads to a steady increase in the central pressure.

As discussed by \citet{2017ApJ...846..170T}, if a circular binary loses an amount of mass $\Delta M$ in an amount of time that is small compared to the orbital period, and there is no kick involved -- i.e., a  symmetric SN -- the post-SN eccentricity 
\begin{equation}
    e = \frac{\Delta M}{M_T},
\end{equation}
where $M_T$ is the total mass of the binary after the mass loss. Given the binding energies of neutron stars, the expected post-SN eccentricities are of the order of 0.1, as observed for the first five eccentric binary MSPs in Table 2.

Although these hypotheses naturally predict not only the observed eccentricities, but also the observed orbital periods, they also predict a rather narrow set of masses for the resulting MSPs, e.g. \citet{2014MNRAS.438L..86F} predict $1.22 M_{\odot} < M_p < 1.27 M_{\odot}$.  \citet{2015ApJ...807...41J} predict larger masses, depending on the unknown threshold for the central density at which the hypothetical phase transition occurs.   Using models with fast rotating neutron stars, they predict a range from $1.4$ to $> 2 \, M_{\odot}$. Such focused MSP masses are incompatible with the mass measurements in Table~\ref{tab:eMSPs}, which span a wide range from 1.35 to $1.83 M_{\odot}$.

Furthermore, such a phase transition is not a likely formation mechanism for \psrb: as mentioned above, the expected eccentricities are too large. The simulations made by \cite{2014MNRAS.438L..86F} suggest that, for any scenario involving a phase transition, fine tuning of mass loss and kick direction and magnitude would be necessary to obtain $e \sim 0.0074$.

An alternative hypothesis, put forward by \cite{2014ApJ...797L..24A}, postulates that the eccentricities of these systems arise from the dynamical interaction of the binary with a circumbinary disk, which results from thermonuclear hydrogen flashes at the surface of the WD. This hypothesis is compatible with the wide range of MSP masses measured for these systems, and it predicts that the eccentric MSP--WD systems have a relatively wide range of orbital periods, from well below 10 d to a maximum of $\sim\,$50 d. The latter value corresponds, according to the relation of \cite{1999A&A...350..928T}, to the critical He WD mass above which hydrogen flashes cease, then thought to be $\sim 0.31 \, M_{\odot}$. More recently, \cite{2021ApJ...909..161H} have suggested that these orbital eccentricities might be the result of a collection of small kicks to the companion WD caused by these thermonuclear flashes. They note that there is currently no consensus on the upper WD mass limit for which hydrogen flashes occur, and suggest that the observed range of orbital periods for the eccentric MSPs could be used for determining this number.

Thus, if either hypothesis is correct, the eccentricity and orbital periods of \psrb\, imply that some amount of thermonuclear activity can still happen for WDs with the mass of the companion of \psrb, which, according to the relation of \cite{1999A&A...350..928T} is $\sim \, 0.32 \, M_{\odot}$.

Very recently, the work of \citet{1992RSPTA.341...39P} has  been expanded by \citet{gc21} in an attempt to explain the high eccentricities of these pulsars. This hypothesis explains the orbital eccentricities in terms of coherent resonances between the orbital period and the timescale of convective eddys in the red giant progenitors, which had already been estimated by \cite{1992RSPTA.341...39P} to be $\sim$25 d. To construct a viable model, Ginzburg \& Chiang had to assume that in resonance - and only in resonance - the convection stops being random. This explains the characteristics of the eccentric MSPs well, especially their orbital periods: the 25-d periodicity lies near the middle of their range of orbital periods. However, it is not yet clear whether this hypothesis can explain the eccentricity of \psrb\, which has a much longer orbital period. Perhaps it is in a weaker 1:2 resonance with the convection turnover time. The orbital period distribution of many more future eccentric MSP discoveries might allow the identification of such multiple resonances.

\subsection{The Galactic MSP population}
\label{sec:GalacticMSPs}

\begin{figure*} 
\includegraphics[width=\textwidth]{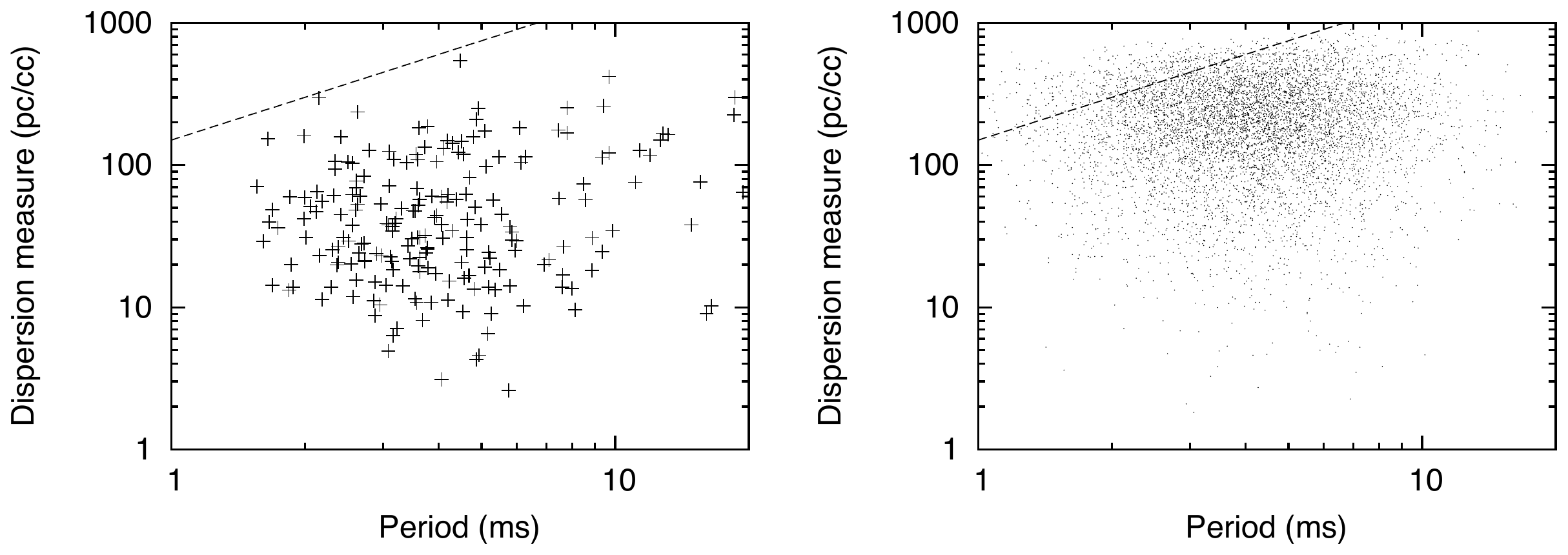}
\caption{Left: DM versus $P$ for the current sample of MSPs. Right: DM versus $P$ for a fake sample of model pulsars in which no selection effects are applied. The dashed line shown in both plots is the locus of points for which DM$/P=150$~cm$^{-3}$~pc~ms$^{-1}$.}
\label{fig:dmp}
\end{figure*}

We conclude with some brief remarks concerning the state of MSP statistics and demography. From the earliest days of this field, when the sample numbered only a few objects, much debate has taken place on the birth rate and population size of MSPs both in the Galaxy and its globular cluster systems. The modest addition of these three pulsars to the observed sample is perhaps notable by the fact that discoveries are still being made of objects which buck the trend based on previous understanding. While population analyses can now make use of a sample that is much larger than in previous years and are strongly encouraged, it is important to realize that these ``outlier'' pulsars such as \psrb\ still being found indicates that the sample is still heavily biased by observational selection. MSPs discovered using the $\gamma$-ray unassociated sources have broadened the selection \citep{Ray2012psc}, but substantial bias remains.

One way to see the limitations of the observed sample is shown in Fig.~\ref{fig:dmp}. In the left panel, we show the observed sample as a scatter diagram in the DM--$P$ plane. Most of the currently known MSPs have DMs in the range 10--100~cm$^{-3}$~pc. The right-hand panel of Fig.~\ref{fig:dmp} shows the results of a simple Monte Carlo simulation and our current model of the underlying MSP population. The Monte Carlo pulsars were generated using the {\tt psrpop} software package \citep{2006MNRAS.372..777L} and was normalized to mimic the sample of 60 MSPs given in \citet{2015MNRAS.450.2185L} and adopting the period distribution derived in that study. For the purposes of this demonstration, we adopt a 500~pc scale height, 10\% duty cycle for intrinsic pulse widths, and take the underlying luminosity distribution to be log-normal, as found for the normal pulsar population \citep{2006MNRAS.372..777L}. It is particularly striking that the mean DM value in the population as a whole is much larger than in the observed sample, and that there is a significant number of pulsars with $ \textrm{DM}/P > 150$~cm$^{-3}$~pc~ms$^{-1}$ that are yet to be discovered.

\section{Conclusions}\label{sec:conc}

We have presented high-precision timing measurements for three Galactic MSPs. One of these is apparently solitary (\psra) while the other two  (\psrc\ and \psrb) are in binary systems. The three timing solutions presented in this paper significantly improve upon what was known about each pulsar prior to this work. The pulsars themselves are generally quite weak and close to the detection threshold available with Parkes. For this reason, the TOA precision is relatively poor, making the derivation of their timing solutions unusually difficult. This required the use of the innovative algorithm described by \citet{2018MNRAS.476.4794F}. In particular, \psrb\, was perhaps the most challenging pulsar ever phase connected with this algorithm; this required the vast improvement in efficiency provided by the implementation of the partial solution prioritisation described in the last paragraph of section 4.3 of \citet{2018MNRAS.476.4794F}. Even with this improvement, the algorithm still needed to analyse 173,311 partial solutions before finding the correct rotation count for \psrb; this took about 8 hours on a single core.

Our work showed that \psrb\ has an anomalously high orbital eccentricity when compared to most low-mass binary MSPs and we speculate that it is a member of a group of unusually eccentric MSP--He WD binaries for which the formation mechanism is not yet well understood. 

Our current timing precision and data span prevent the measurement of proper motions, parallaxes and post-Keplerian effects such
as the rate of advance of periastron ($\dot{\omega}$) or the Shapiro delay. Some of these effects will be measurable with MeerKAT \citep{MeerKATpsrs} observations: \psrc\, and \psra\, were observed by Spiewak et al. (in prep) and seen with S/N $>50$ in $<30$ minutes. The dense MeerKAT timing campaigns carried out by the relativistic binaries program \citep{2021MNRAS.504.2094K} will be able to measure the $\dot{\omega}$ of \psrb\, to high significance, and thus determine its total mass, although this will likely need a couple of orbital campaigns spaced by a few years. If the low orbital inclinations inferred above for \psrc\, and \psrb\, are correct, then a detection of the Shapiro delay will be difficult for either system, even with dense MeerKAT campaigns, and will likely have to wait for instruments like the Square Kilometre Array. Long-term, low-cadence observations will measure the proper motions and increase the sensitivity to $\gamma$-ray pulsations for the three pulsars. Such measurements could reveal additional companions to these pulsars that are not currently detectable \citep[e.g., as for PSR~J1024$-$0719,][]{2016ApJ...826...86K,2016MNRAS.460.2207B}.

Finally, we highlight that the MSP sample is incomplete and predict that further unusual discoveries will be made as we better sample the Galactic MSP zoo. Currently, these rare moderately eccentric MSPs allow a glimpse into the final evolution of mass transfer/mass loss in binary pulsars. As we discover more, the bulk properties of the population will enable us to determine at 
what orbital separations and donor masses episodic mass loss may be important.

\section*{Data availability}

The data underlying this article will be shared on reasonable request to the corresponding author.

\section*{Acknowledgements}

We thank Paul Ray, Ben Perera and the anonymous referee for useful comments on an earlier version of this manuscript. DRL and MAM acknowledge support from the NSF awards AAG-1616042, OIA-1458952 and PHY-1430284. 

Work at NRL is supported by NASA. We acknowledge the Wiradjuri people as the traditional owners of the Observatory site. The Parkes telescope (``Murriyang'') is part of the Australia Telescope National Facility which is funded by the Commonwealth of Australia for operation as a National Facility managed by CSIRO.  

The \textit{Fermi} LAT Collaboration acknowledges generous ongoing support
from a number of agencies and institutes that have supported both the
development and the operation of the LAT as well as scientific data analysis. These include the National Aeronautics and Space Administration and the Department of Energy in the United States, the Commissariat \`a l'Energie Atomique and the Centre National de la Recherche Scientifique / Institut National de Physique Nucl\'eaire et de Physique des Particules in France, the Agenzia Spaziale Italiana and the Istituto Nazionale di Fisica Nucleare in Italy, the Ministry of Education, Culture, Sports, Science and Technology (MEXT), High Energy Accelerator Research Organization (KEK) and Japan Aerospace Exploration Agency (JAXA) in Japan, and the K.~A.~Wallenberg Foundation, the Swedish Research Council and the Swedish National Space Board in Sweden.
 
Additional support for science analysis during the operations phase is gratefully acknowledged from the Istituto Nazionale di Astrofisica in Italy and the Centre National d'\'Etudes Spatiales in France. This work performed in part under DOE
Contract DE-AC02-76SF00515.

\bibliographystyle{mnras}

\bsp	
\label{lastpage}
\end{document}